\documentclass[showpacs,preprintnumbers,preprint,nofootinbib]{revtex4}
\usepackage{amssymb}
\usepackage{amsmath}
\usepackage{graphicx}
\usepackage{dcolumn}
\usepackage{bm}
\usepackage{epsf}

\def\be{\begin{eqnarray}}
\def\ee{\end{eqnarray}}
\def\bea{\begin{eqnarray}}
\def\eea{\end{eqnarray}}

\def\bm{\boldsymbol}

\def\vR{{\bm R}}

\def\vj{{\bm j}}

\def\vr{{\bm r}}
\def\vq{{\bm q}}
\def\vk{{\bm k}}
\def\vp{{\bm p}}

\def\vx{{\bm x}}
\def\vy{{\bm y}}

\def\vs{{\bm \sigma}}

\def\vtau{{\bm \tau}}

\def\del{{\partial}}
\def\hatr{{\hat \vr}}

\newcommand{\no}{\nonumber \\}
\newcommand{\etal}{{\it et al.}~}

\def\H#1{{}^{#1}\mbox{H}}
\def\He#1{{}^{#1}\mbox{He}}
\def\nlo#1{\mbox{N$^{#1}$LO}}

\begin{document}

\preprint{HEP/123-qed}
\title{
Up-to N$^3$LO
heavy-baryon chiral perturbation theory calculation
for the M1 properties of three-nucleon systems
}
\author{Young-Ho Song}
\email{yhsong@phy.duke.edu}
\affiliation{Department of Physics, Duke University,
Durham, NC 27708, USA}
\author{Rimantas Lazauskas}
\email{rimantas.lazauskas@ires.in2p3.fr}
\affiliation{IPHC, IN2P3-CNRS/Universit\'e Louis Pasteur BP 28, F-67037 Strasbourg Cedex
2, France}
\author{Tae-Sun Park}
\email{tspark@kias.re.kr}
\affiliation{
Department of Physics and BAERI,
Sungkyunkwan University,
Suwon 440-746, Korea}
\date{\today }

\begin{abstract}
M1 properties, comprising magnetic moments and radiative capture of thermal
neutron observables, are studied in two- and three-nucleon systems. We utilize
meson exchange current derived up to N$^3$LO using heavy baryon chiral
perturbation theory a la Weinberg. Calculations have been performed for
several qualitatively different realistic nuclear Hamiltonians, which permits
us to analyze model dependence of our results.
Our results are found to be strongly correlated with the effective range parameters
such as binding energies and the scattering lengths.
Taking into account such correlation, the results are in good agreement with the
experimental data with small model-dependence.
\end{abstract}

\maketitle

\renewcommand{\thefootnote}{\#\arabic{footnote}}
\setcounter{footnote}{0}

\section{Introduction}

M1 properties -- nuclear magnetic moments as well as radiative capture cross sections --
are the fundamental low-energy observables of a few nucleon systems
and therefore presents ideal laboratory to test effective field theories (EFTs).
In this regard, M1 properties have been extensively studied using EFT
with huge successes~\cite{npdg,M1S,npAsym,chen,sadeghi,skibinski,A23}.
%finding an huge success
One such example is the ability to describe $\sigma_{np}$,
the capture cross section of the $np\to d\gamma$ process,
at threshold with 1 \% accuracy
by applying heavy-baryon chiral perturbation theory (HBChPT)
up to next-next-next-to the leading order or \nlo3~\cite{npdg}.
In this work, we extend our up-to \nlo3 HBChPT description
of the M1 properties to $A=3$ systems.
By taking the magnetic moments of $\H3$ and $\He3$ as input
to fix the coefficients of the contact-term operators,
a completely parameter-free theory predictions will be made
for the total cross section and the photon polarization of
the  thermal neutron capture process ($nd\to \H3\gamma$).
We will also revisit the theory predictions
for the two-body observables:
deuteron magnetic moment $\mu_d$ and $\sigma_{np}$.

The purpose of this article is to demonstrate the general tenet of EFTs
by studying the M1 properties of a few body systems:
once the long-range contributions are taken into account correctly,
EFTs enables accurate and model-independent results,
regardless the details of the short-range
%(or high-energy)
physics.
% provided that a proper renormalization has
%been performed.

%To explain the above point,
%it might be helpful to
We begin with a few comments that are generic to all EFTs.
At a certain order in EFTs, there appear
%counter-terms or
contact-terms (CTs),
which parameterize the high-energy (or short-range) physics
above the cutoff scale of the theory.
The coefficients of CTs -- which we refer to as low-energy constants (LECs) --
are thus sensitive to the short-range physics, and depend on
the adopted cutoff value and the regularization/renormalization scheme.
The values of LECs are not fixed by the symmetry alone,
and should be determined by either solving the underlying theory
or by fitting them
so as to
%in order to
reproduce selected set of known experimental data.
Since the former is currently not feasible,
the latter remains the only practical option.
At \nlo3, HBChPT M1 currents contain two
non-derivative
two-nucleon CTs,
one in iso-vector and the other in iso-scalar channel.
%The LECs of them,
These LECs, $g_{4v}$ and $g_{4s}$,
will be determined in this work
by requiring to reproduce the experimental values
of
%$\mu(\H3)$ and $\mu(\He3)$,
the magnetic moments of $\H3$ and $\He3$.
Once these LECs fixed,
we are left with no free parameters
and can make totally parameter-free
theory predictions for the other M1 observables\footnote{
There are many other alternatives.
For example, one can fix $g_{4v}$ and $g_{4s}$ from the experimental
values of $\sigma_{np}$ and the deuteron magnetic moment $\mu_d$,
and then make theory predictions on $\mu(\H3)$ and $\mu(\He3)$~\cite{A23}.
}.
Note that the CT contributions have been ignored
in Ref.~\cite{npdg}, which caused
%small but noticeable
small cutoff-dependence in $\sigma_{np}$.
By taking into account of the LECs,
we will show that $\sigma_{np}$ becomes virtually completely cutoff-independent.
Second comment is about the accuracy of the adopted wave functions at short range.
The authors of Ref.~\cite{MEEFT}
have developed an approach
called EFT$^*$ or
MEEFT ({\em more effective} effective field theory)
that enables a consistent and systematic
EFT calculations on top of accurate but phenomenological wave functions.
The key observations are following.
The model-dependence resides mainly in the short-range region of the wave functions.
Since short-range contributions can be well embodied by local operators at low-energy
and since EFT has the machinery to contain all the relevant
local operators (i.e., CTs) in a consistent and systematic manner,
the model-dependence due to short-range physics is to be absorbed into the
renormalization procedure of the LECs.
%As a result, the values of the LECs are model-dependent, while the net results are
%model-independent.
To be more specific,
if we adopt other wave functions that have different short-range behavior,
the values of LECs should also be changed so as to reproduce the selected
experimental data {\em with} the adopted wave functions.
By performing this procedure,
while the values of LECs -- which
are not physical observables  --
are model-dependent, the resulting net contributions become model-independent.
An easy and effective way of proving the model-independence
in a quantitative fashion might be to look at
the cutoff-dependence of the results,
since the cutoff value is the key parameter that characterize the short-range contributions.
Such a numerical proof will be taken in this work.
The third comment is about the long-range contributions.
Note that mismatches in the long-range contributions cannot be cured by
finite set of local operators.
The long-range part of the transition operator
is usually governed by the chiral symmetry,
leaving little uncertainty there.
On the other hand,
the long-range part of the wave functions is controlled by the
effective range parameters (ERPs) such as
the nuclear binding energies and the scattering lengths.
For two-nucleon systems, most of the modern realistic $NN$ potentials
reproduce the ERPs with a great accuracy.
%However for three and more nucleon systems situation becomes highly non-trivial
However for nucleon systems with $A\ge 3$,
situation becomes highly non-trivial
as many of the available potentials fail to reproduce
the relevant ERPs to the desired accuracy.
%simultaneously.

As we will demonstrate, our results have little cutoff-dependence for all the cases considered,
which might be interpreted that the short-range physics is
well under control.
% using LECs.
On the other hand,
the model-dependence due to the difference in long-range part of the wave functions  will
cause correlations of the matrix elements with the ERPs.
%, since ERPs govern the
%long-range behavior.
%
In our work, we observe rather a strong model-dependence
and demonstrate how it is correlated to the
%closely related with the
model prediction of the triton binding energy $B_3$.
It indicates that the model-dependence is
due to the mismatches in the long-range contributions.

To bypass the difficulty and to get model-independent accurate theory
predictions, we have explored two different approaches.
One is to
bring prediction of $B_3$ to its experimental value $B_3^{\rm exp}= 8.482$ MeV,
using the observed correlation curves.
The resulting M1 matrix elements are found to be model-independent to a good accuracy,
and consistent with the experimental data.
Another way is
to adjust the tri-nucleon interactions (TNIs) to meet
the experimental values of the ERPs.

\section{Formalism}

\subsection{Faddeev\ equations \label{sec_FY_eq}}

During the last few decades several different methods permitting to solve
three body bound and scattering problem has been developed. In this study we
solve Faddeev~\cite{Fadd_art} equations (also often called Kowalski-Noyes
equations) in configuration space to obtain 3-body bound and scattering wave
functions. We employ the isospin formalism, i.e.,
consider proton and neutron
as two degenerate states of the same particle - nucleon, having the mass
fixed to $\hbar ^{2}/m=41.471$ MeV$\cdot$fm. Then three Faddeev equations
become formally identical, having the form
\begin{equation}
\left( E-H_{0}-V_{ij}\right) \Phi _{ij,k}=V_{ij}(\Phi _{jk,i}+\Phi _{ki,j}),
\label{EQ_FE}
\end{equation}
where $(ijk)$ are particle indices, $H_{0}$ is kinetic energy operator,
$V_{ij}$ is two body force between particles $i$ and $j$, $\Phi _{ij,k} $ is
Faddeev component.
It is useful to define cyclic ($P^{+}$) and anti-cyclic
($P^{-}$) particle permutation operators, which permits to transform Faddeev
component between two particle bases: $P^{+}=(P^{-})^{-1}=P_{23}P_{12}$ and
$P^{+}\Phi _{ij,k}=\Phi _{jk,i},$ while $P^{-}\Phi _{ij,k}=\Phi _{ki,j}$.
%Systems wave function in Faddeev formalism is the sum of three Faddeev
The wave function in Faddeev formalism is the sum of three Faddeev
components, which employing permutation operators can be written as:
\begin{equation}
\Psi =(1+P^{+}+P^{-})\Phi _{ij,k} .  \label{EQ_WF}
\end{equation}

Faddeev components, if represented in its proper coordinate basis, have
simple structure and analytical asymptotic behavior for the short-range potentials.
We use relative Jacobi coordinates
$\vx_{k}=(\vr_{j}-\vr_{i})\smallskip $
and
$\vy_{k}=
\frac{2}{\sqrt{3}}(\vr_{k}-\frac{\vr_{i}+\vr_{j}}{2})$,
whereas Faddeev components we expand in bipolar harmonic basis:
\begin{equation}
\Phi _{ij,k}=\sum\limits_{\alpha }\frac{F_{\alpha }(x_{k},y_{k})}{x_{k}y_{k}}%
\left\vert \left( l_{x}\left( s_{i}s_{j}\right) _{s_{x}}\right)
_{j_{x}}\left( l_{y}s_{k}\right) _{j_{y}}\right\rangle _{JM}\otimes
\left\vert \left( t_{i}t_{j}\right) _{t_{x}}t_{k}\right\rangle _{TT_{z}},
\label{EQ_FA_exp}
\end{equation}%
here index $\alpha $ represents all the symmetry allowed combinations of the
quantum numbers presented in the brackets: $l_{x}$ and $l_{y}$ are the
partial angular momenta associated with respective Jacobi coordinates; $%
s_{i} $ and $t_{i}$ are the spins and isospins of the individual particles. Functionals
$F_{\alpha }(x_{k},y_{k})$ are called partial Faddeev amplitudes.
Three-nucleon system conserves its total angular momentum $J$ as well as its
projection $M$, however due to the presence of charge dependent terms in nuclear
interaction, total isospin of the system $T$ is not conserved.

Equation~(\ref{EQ_FE}) is not complete, it should be complemented with the
appropriate boundary conditions. Boundary conditions can be written in the
Dirichlet form. First Faddeev amplitudes, for bound as well as for
scattering states, satisfy the regularity conditions:
\begin{equation}
F_{\alpha }(0,y_{k})=F_{\alpha }(x_{k},0)=0.  \label{BC_xyz_0}
\end{equation}%
For the bound state problem
%systems
wave function is compact, therefore the
regularity conditions can be completed by forcing the amplitudes
$F_{\alpha} $
to vanish at the borders of a hypercube
$\left[ 0,X_{\max }\right] \times \left[ 0,Y_{\max }\right] $:
\begin{equation}
F_{\alpha }(X_{\max },y_{k})=F_{\alpha }(x_{k},Y_{\max })=0.
\end{equation}%
Finally, we normalize three-nucleon wave function to unity
$\left\langle \Psi \mid \Psi \right\rangle =1.$

%\bigskip

Faddeev components describing neutron-deuteron scattering, for the energies below the break-up threshold,
 vanish  for $\mathbf{x}_{k}\rightarrow\infty$.
As $\mathbf{y}_{k}\rightarrow\infty $ interaction between particle $k$ and cluster $ij$
is negligible and Faddeev components $\Phi _{jk,i}$ and $\Phi _{ki,j}$ vanish. Then
the component $\Phi _{ij,k}$ describes the plane wave of the particle $k$ with respect
to the bound particle pair $ij$:
\begin{eqnarray}
\lim_{y_k\to \infty}
\Phi_{ij,k}(\mathbf{x}_{k},\mathbf{y}_{k} )&=&
\frac{1}{\sqrt{3}}\sum\limits_{j_{n}^{\prime }l_{n}^{\prime }}
\left| \left\{\psi_{d}(\mathbf{x}_{k})\right\}_{j_{d}}\otimes \left\{ Y_{l_{n}}(\mathbf{\hat{y}}_{k})\otimes s_{k}\right\}
_{j_{n}}\right\rangle_{JM}
%
%\otimes \left\vert \left( t_{i}t_{j}\right)_{t_{d}}t_{k}\right\rangle _{(1/2)(-1/2)}
\otimes \left\vert \left( t_{i}t_{j}\right)_{t_{d}}t_{k}\right\rangle_{\frac12,-\frac12}
\notag \\
&&
\times \frac{i}{2}\left[ h_{l_{n}}^{-}(pr_{nd})-S_{j_{n}l_{n},j_{n}^{%
\prime }l_{n}^{\prime }}h_{l_{n}}^{+}(pr_{nd})\right],  \label{eq_as_beh}
\end{eqnarray}%
where deuteron, being formed from nucleons $i$ and $j$, has quantum numbers
$s_{d}=1$,  $j_{d}=1$ and $t_{d}=0$ and its wave function
$\psi _{d}(\mathbf{x}_{k})$ is normalized to unity;
$p$ designates the relative momentum of
incoming neutron, $r_{nd}=(\sqrt{3}/2)y_{k}$ is relative distance between
neutron and deuteron target, whereas $h_{l_{n}}^{\pm }$ are
%well known
the spherical Hankel
functions. Expression~(\ref{eq_as_beh}) is normalized so that
$nd$ scattering
wave function has unity flux.

For zero or very low momentum neutrons,
as is the case for the thermal neutron capture,
only relative $S$-wave amplitudes survives in the asymptote, whereas
expression~(\ref{eq_as_beh}) simplifies to:
\begin{eqnarray}
\lim_{y_k\to \infty}
\Phi_{ij,k}(\mathbf{x}_{k},\mathbf{y}_{k} )&=&
\frac{1}{\sqrt{3}}\sum\limits_{j_{n}^{\prime }l_{n}^{\prime }}
\left| \left\{\psi_{d}(\mathbf{x}_{k})\right\}_{j_{d}}\otimes \left\{ Y_{l_{n}}(\mathbf{\hat{y}}_{k})\otimes s_{k}\right\}
_{j_{n}}\right\rangle_{JM}
%
%\otimes \left\vert \left( t_{i}t_{j}\right)_{t_{d}}t_{k}\right\rangle _{(1/2)(-1/2)}
\otimes \left\vert \left( t_{i}t_{j}\right)_{t_{d}}t_{k}\right\rangle_{\frac12,-\frac12}
\notag \\
&&
\times \left[ 1-\frac{^{2J+1}a_{nd\text{ }}}{r_{nd}}\right] ,
\label{eq_as_beh0}
%
%\Phi _{ij,k}(\mathbf{x}_{k}\mathbf{,y}_{k} &\rightarrow &\infty )=\frac{1}{%
%\sqrt{3}}\sum\limits_{j_{n}^{\prime }l_{n}^{\prime }}\left. \left\{ \psi
%_{d}(\mathbf{x}_{k})\otimes \left( s_{i}s_{j}\right) _{s_{d}}\right\}
%_{j_{d}}\otimes \left\{ Y_{l_{n}}(\mathbf{y}_{k})\otimes s_{k}\right\}
%_{j_{n}}\right\rangle _{JM}\otimes \left\vert \left( t_{i}t_{j}\right)
%_{t_{d}}t_{k}\right\rangle _{(1/2)(-1/2)} \\ \notag
%&&\times \left[ 1-\frac{^{2J+1}a_{nd\text{ }}}{r_{nd}}\right] ,
\end{eqnarray}%
where $^{2J+1}a_{nd}$ is neutron-deuteron scattering length.
For the cases where Urbana type three-nucleon interaction (TNI) are included,
noting that the TNI among particles $ijk$ can be written as sum of three terms
$V_{ijk}=V_{ij}^{k}+V_{jk}^{i}+V_{ki}^{j}$,
we modify the Faddeev equation (\ref{EQ_FE}) into:
\begin{equation}
\left( E-H_{0}-V_{ij}\right) \Phi _{ij,k}=V_{ij}(P^{+}+P^{-})\Phi_{ij,k}+\frac{1}{2}(V_{jk}^{i}+V_{ki}^{j})\Psi .
\end{equation}%
%where the three-nucleon force
%and the resulting Faddeev equation reads
%Original Faddeev equation is formulated for the particles interacting only
%by short range two-body forces. In this paper we will also present results
%for the Hamiltonians, which contain Urbana type three-body interactions. In this aim, we
%modify Faddeev equation in order to include 3NF:
%where three-nucleon force between particles $ijk$ is a sum of three terms
%$V_{ijk}=V_{ij}^{k}+V_{jk}^{i}+V_{ki}^{j}$.

\subsection{Electromagnetic current}

For three-body system one has three 1-body currents associated with each
particle and three 2-body currents associated with each pair of particles.
Thus
\begin{equation}
J_{em}=\sum\limits_{i=1,i\neq (j<k)}^{3}(J_{1B}^{(i)}+J_{2B}^{(jk)}).
\end{equation}
%
%Three-nucleon wave-functions $\left\vert \Psi \right\rangle $ in isospin
%formalism is fully antisymmetric, therefore:
Since the wave-functions $\left\vert \Psi \right\rangle $ in isospin
formalism is fully antisymmetric, the matrix element of the current operators
can be written as
\begin{equation}
\left\langle \Psi _{f}\right\vert J_{em}\left\vert \Psi _{i}\right\rangle
=\sum\limits_{i=1,i\neq (j<k)}^{3}\left\langle \Psi _{f}\right\vert
J_{1B}^{(i)}+J_{2B}^{(jk)}\left\vert \Psi _{i}\right\rangle =3\left\langle
\Psi _{f}\right\vert J_{1B}^{(3)}\left\vert \Psi _{i}\right\rangle
+3\left\langle \Psi _{f}\right\vert J_{2B}^{(12)}\left\vert \Psi
_{i}\right\rangle,  \label{eq_3x_rule}
\end{equation}

We use the electromagnetic current operators derived from HBchPT,
which contain the nucleons and pions as pertinent degrees of freedom with all other
massive fields integrated out.
In HBchPT the electromagnetic currents and M1 operator are expanded systematically with increasing
powers of $Q/\Lambda_\chi$ , where $Q$ stands for the typical momentum scale of the process
and/or the pion mass, and $\Lambda_\chi \sim 4\pi f_\pi \sim m \sim 1$ GeV is the chiral scale,
$f_\pi\sim 92.4$ MeV is the pion decay constant,
and $m$ is the nucleon mass. We remark that, while the nucleon
momentum $\vp$ is of order of $Q$, its energy $(\sim\frac{\vp^2}{m})$ is of order of $Q^2/m$,
and consequently the four-momentum of the emitted photon $q^\mu=(\omega,\,\vq)$
with $|\vq| = \omega$
also is counted as $O(Q^2/m)$.
%We restricted our calculation of
Current operators are obtained up to \nlo3.
%$N^3 LO$ in chiral order counting because
Note that three-body currents are \nlo4 or higher order,
and do not enter in our work.~\footnote{
It is worth mentioning that there is a
different power counting scheme where the nucleon mass is regarded as heavier than the chiral scale,
$m \sim \Lambda_\chi^2/Q$,
see refs.~\cite{POT_BN3LO,EGM_NPA671_00} for details.
However, the use of this alternative
counting scheme would not affect the results to be reported in this work since the difference
between the two counting schemes would appear only at higher orders than
explicitly considered here (\nlo3).
}

Let us list the relevant current operators. The explicit form of magnetic moment operators
can be found in the ref.~\cite{A23}.
The one-body current including the relativistic corrections reads
\bea
J_{1B}^{(i)}(\vq;\vr_i)=&e^{-i\vq\cdot \vr_i}
&\bigg[\frac{Q_i}{m}\bar{\vp}_i\left(1-\frac{\bar{\vp}_i^2 }{2m}\right)
 +\frac{1}{2m}i\vq\times\vs_i\left(\mu_i-\frac{Q_i}{2m^2}\bar{\vp}_i^2 \right) \no
&                       &-\frac{\omega (2\mu_i-Q_i)}{8m^2}(2i\bar{\vp}_i\times\vs_i)
                      -\frac{\mu_i-Q_i}{16 m^3}
                     (4i\vq\times \bar{\vp}_i\vs_i\cdot \bar{\vp}_i)\no
%&                       &+(\mbox{longitudinal parts})
& &-\frac{w(2\mu_i-Q_i)}{8m^2} \vq-\frac{\mu_i-Q_i}{16 m^3}(-2\vq \vq\cdot\bar{\vp}_i)
                                         +(\mbox{higher orders})\bigg]
\eea
where $Q_i$ and $\mu_i$ represent the charge and magnetic moment of $i$-th nucleon,
and $\bar{\vp}\equiv \frac12 (i\stackrel{\leftarrow}{\nabla}
-i\stackrel{\rightarrow}{\nabla})$ should be understood to act only on the
nuclear wave functions.

\begin{figure}
\begin{center}
\label{fig:1pi}
\includegraphics[]{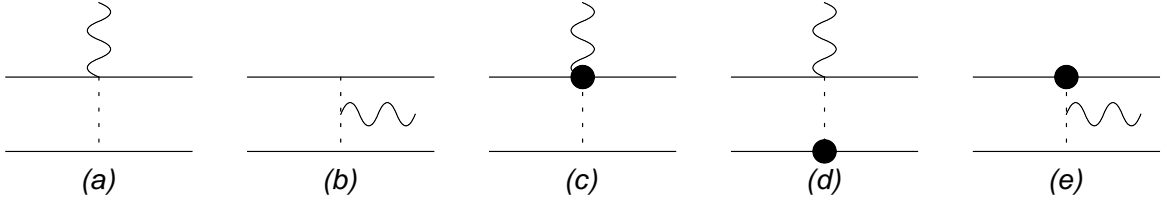}
\caption{Tree diagrams for the electromagnetic current operators.
%graphs of NLO and \nlo3.
Soft one-pion-exchange,
the sum of the ``seagull"$(a)$ and the ``pion-pole" $(b)$ diagrams
contribute to the $J_{1\pi}$.
Diagrams $(c)-(e)$ contribute to the
$J_{1\pi C}$ at \nlo3.
The dot represents
the vertex corrections coming from NLO or \nlo2 lagrangian.}
\end{center}
\end{figure}

Corrections to the 1B operator are due to the meson-exchange currents (MECs).
Up to \nlo3, as mentioned,
only two-body (2B) contributions enter.
% three-body (3B) currents are \nlo4 or higher order.
It is to be emphasized that MECs derived in EFT are meaningful
only up to a certain momentum scale characterized by the cutoff $\Lambda$.
%, where $\Lambda$ is the cutoff below which the chosen explicit degrees of freedom reside.
In our work,
we adopt a Gaussian regulator
in performing the Fourier transformation of the MECs from momentum
space to coordinate space~\cite{MEEFT}.
It is to be noted that the contributions due to high momentum exchanges (above the cutoff scale) are
not simply ignored but, as we will discuss later, they are accounted for by the renormalization of the contact-term coefficients.

We decompose the two-body current into the soft-one-pion-exchange ($1\pi$),
vertex corrections to the one-pion exchange $(1\pi C)$, the two-pion-exchanges $(2\pi)$,
and the contact-term ($CT$)contributions,
\be
J_{2B}^{(jk)}&=&{J}^{(jk)}_{1\pi}+{J}^{(jk)}_{1\pi C}+{J}^{(jk)}_{2\pi}+{J}^{(jk)}_{CT}.
\ee
It is noteworthy that
there can be additional corrections to the 2-body current coming from the so-called fixed term.
The ¡°fixed-term¡± contributions represent vertex corrections to the soft-one-pion-exchange
and fixed completely by Lorentz covariance.
Because the fixed terms make the calculation highly involved, but only give
very small contributions in M1 operator according to our previous study~\cite{A23},
we neglected the fixed term contributions in the present work.
%Because the fixed terms, containing the nucleon momentum operators, make the calculation highly involved, we will
%not include those contributions in the present work. In our previous study~\cite{A23} however we have demonstrated
%that the fixed term contributions in M1 operator are very small.

The soft-one-pion exchange current ${J}^{(jk)}_{1\pi}$ is NLO and can be written
in terms of $\vR_{jk}=\frac{1}{2}(\vr_j+\vr_k)$, $\vr=\vr_j-\vr_k$, ${\hat \vr}=\vr/|\vr|$,
$S_{jk}=3\vs_j\cdot{\hat \vr}\,\vs_k\cdot{\hat \vr}-\vs_j\cdot\vs_k$,
\bea
\label{j:OPE:r}
{J}_{1\pi}^{(jk)}(\vr,\vR)&=&e^{-i\vq\cdot\vR}
       \bigg\{-\frac{g_A^2 m_\pi^2}{12 f_\pi^2}({\vec\tau}_j\times{\vec\tau}_k)^z \vr
          \left[{\vec\sigma}_j\cdot{\vec\sigma}_k \left(y^\pi_{0\Lambda}(r)-\frac{\delta_\Lambda(r)}{m_\pi^2}\right)
          +S_{jk}y_{2\Lambda}^\pi(r)\right] \no
        &&+i\frac{g_A^2}{8f_\pi^2}\vq\times
           \left[{\hat T}_{S,jk}^{(\times)}\left(\frac{2}{3}y_{1\Lambda}^\pi(r)
           -y_{0\Lambda}^\pi(r)\right)
                    -{\hat T}_{T,jk}^{(\times)}y_{1\Lambda}^\pi(r)\right]\bigg\},\nonumber
\eea
where
%\footnote{We neglected small corrections to the ${J}_{1\pi}$
%expression which comes from Fourier transformation
%from momentum space to configuration space. We have
%verified they are numerically negligible~\cite{A23}.
%Strictly speaking, Fourier transformation of ${J}_{1\pi}$
%from momentum space to configuration space can have $N^3LO$
%and higher order corrections. We have verified however that
%$N^3 LO$ corrections to the soft one pion exchange current are
%numerically negligible.
%}
\bea
\label{j:hatTs}
{\hat T}_{S,jk}^{(\odot)} &=& (\tau_j \odot \tau_k)^z
({\bm \sigma}_j \odot {\bm \sigma}_k),
 \nonumber \\
{\hat T}_{T,jk}^{(\odot)} &=& (\tau_j \odot \tau_k)^z \left[{\hat
r}\,{\hat r}\cdot ({\bm \sigma}_j \odot {\bm \sigma}_k) -\frac13
({\bm \sigma}_j \odot {\bm \sigma}_k)\right],
\eea
$\odot= \pm,\, \times$, and the regulated delta and Yukawa functions are defined as
\be
\delta_\Lambda(r)&\equiv&\int
 \frac{d^3\vk}{(2\pi)^3} e^{-k^2/\Lambda^2} e^{i\vk\cdot\vr}\no
y_{0\Lambda}^\pi(r)&\equiv&\int
 \frac{d^3\vk}{(2\pi)^3} e^{-k^2/\Lambda^2} e^{i\vk\cdot\vr}
 \frac{1}{\vk^2+m_\pi^2}\no
y_{1\Lambda}^\pi(r)&\equiv&-r\frac{\del}{\del r}y_{0\Lambda}(r), \quad
y_{2\Lambda}^\pi(r)\equiv\frac{r}{m_\pi^2}\frac{\del}{\del r}
 \frac{1}{r}\frac{\del}{\del r} y_{0\Lambda}(r).
\ee

%Among the \nlo3 contributions,
The one-loop vertex correction to the one-pion exchange has been investigated in detail in refs.~\cite{npdg,M1S},
\bea
&{J}^{(12)}_{\rm 1\pi C}=e^{-i\vq\cdot\vR} i \vq \times \bigg\{
-\frac{g_A^2}{8 f_\pi^2}({\bar c}_\omega+{\bar c}_\Delta)
  \big[({\hat T}_{S}^{(+)}+ {\hat T}_{S}^{(-)}) \frac{{\bar y}^\pi_{0\Lambda}}{3}
  +({\hat T}_{T}^{(+)} + {\hat T}_{T}^{(-)})\,y^\pi_{2\Lambda}\big]\no
&+\frac{g_A^2 }{8 f_\pi^2} {\bar c}_{\Delta}
 [\frac13{\hat T}_S^{(\times)}{\bar y}^\pi_{0\Lambda}-\frac{1}{2}{\hat T}_T^{(\times)} y^\pi_{2\Lambda}]
\no
&-\frac{1}{16 f_\pi^2}{\bar N}_{WZ}\vtau_1\cdot\vtau_2
  \big[(\vs_1+\vs_2){\bar y}^\pi_{0\Lambda}
+(3{\hat r}{\hat r}\cdot(\vs_1+\vs_2)-(\vs_1+\vs_2)) y^\pi_{2\Lambda}\big]\bigg\},
\label{j:1piC:r}\eea

The values of the LECs $({\bar c}_\omega,{\bar c}_\Delta,{\bar N}_{WZ})$
should in principle be fixed either by solving the underlying theory, QCD, or by fitting to
suitable experimental observables. Since this has not yet been done, we adopt here the estimates given in refs.~\cite{npdg,M1S}
based on the resonance saturation assumption and the Wess-Zumino action,
$({\bar c}_\omega,{\bar c}_\Delta,{\bar N}_{WZ})\simeq (0.1021, 0.1667, 0.02395)$.

\begin{figure}
\begin{center}
\label{fig:2pi}
\includegraphics[width=8.25cm]{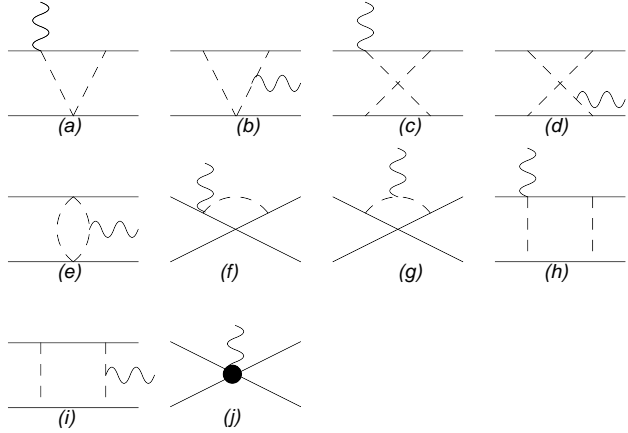}
\caption{Diagrams which contribute to $J_{2\pi}$ $(a)$-$(i)$ and $J_{CT}$ $(j)$ at \nlo3.}
\end{center}
\end{figure}

The two-pion exchange diagrams give rise to
\be
J_{2\pi}^{jk}&=& \frac{e^{-i\vq\cdot\vR}}{128\pi^2 f_\pi^4}\left(
                  i\vq\times[({\hat T}_S^{(+)}-{\hat T}_S^{-})L_S(r)+({\hat T}_T^{(+)}-{\hat T}_T^{-})L_T(r)]
                  -(\tau_j\times\tau_k)^z {\hat r}\frac{d}{dr}L_0(r)\right)\no
\ee
where
\be
L_S(r)&=&-\frac{g_A^2}{3} r\frac{d}{dr} K_0+\frac{g_A^4}{3}(-2 K_0+4 K_1
                + r \frac{d}{d r} K_0 + 2 r \frac{d}{d r} K_1) \no
L_T(r)&=&\frac{g_A^2}{2} r\frac{d}{dr} K_0+\frac{g_A^4}{2}(4 K_{\rm T} - r \frac{d}{d r} K_0
               - 2 r \frac{d}{d r} K_1)\no
L_0(r)&=& 2K_2+g_A^2(8 K_2+2K_1+2K_0)-g_A^4(16 K_2+5K_1+5K_0)+g_A^4\frac{d}{dr}(r K_1),
\ee
and the loop functions $K$'s are defined in refs.~\cite{npdg,MEEFT}.

Finally, contact-term contributions have the form
\be
{J}^{(jk)}_{CT}=e^{-i\vq\cdot\vR}\frac{i}{2m_p}
\vq\times[g_{4S}(\vs_j+\vs_k)
         +g_{4V}T_S^{(\times)}]\delta_\Lambda(r)
\label{jCT}
\ee
where $g_{4S} = m_p g_4$ and $g_{4V} =-m_p(G^R_A+\frac{1}{4} E^{V,R}_T)$.
We remark that, three contact terms were introduced in refs.~\cite{npdg,M1S},
whose coefficients are denoted as $g_4$, $G^R_A$ and $E^{V,R}_T$.
% are the coefficients of the contact terms introduced . A noteworthy point is that, after removing redundant terms,
%
However, due to Fermi-Dirac statistics, only two of them are independent,
and consistent with Eq.~(\ref{jCT}).
A similar
reduction has been noticed for the Gamow-Teller operator,
where only one linear combination of two CTs is required~\cite{MEEFT}.
%LECs needs to be retained.

\subsection{$n$-$d$ radiative capture \label{sec_FY_eq copy(2)}}
In center of mass frame, each currents can be written in the form of
\be
J_{em}&=&e^{-i{\vq}\cdot{\vx}}(i{\vq}\times{\vj}_\mu+{\vj}_c)
\ee
where ${\vx}$ is $\vr_i$ for
$J_{1B}^{(i)}$ and $\vR_{jk}$ for $J_{2B}^{(jk)}$.

To calculate neutron radiative capture observables we will use multipole
expansion. First we introduce a shorthand notations for the multipoles:
\begin{eqnarray}
F_{JM}(\hatr) &=&j_{J}(qr)Y_{JM}(\hatr), \notag \\
F_{JL}^{M}(\hatr) &=&j_{L}(qr)\mathcal{Y}_{JL1}^{M}(\hatr),
\end{eqnarray}
where $j_{L}(qr)$ is
%Ricatti-
the spherical Bessel function; $Y_{JM}$ and $\mathcal{Y}_{JL1}^{M}$ are
spherical and vector-spherical harmonics respectively; $r$ is a vector describing the particle (nucleon or meson), which
interacts with EM field.
%Then for electric and magnetic multipoles one has
%respectively:
Then the electric and magnetic multipoles read
\begin{eqnarray}
\mathcal{M}_{JM} &=&F_{JJ}^{M}(\hatr)\cdot \vj_{c}+iq%
\left[ \left( \frac{J+1}{2J+1}\right) ^{1/2}F_{JJ-1}^{M}(\hatr)-\left(
\frac{J}{2J+1}\right) ^{1/2}F_{JJ+1}^{M}(\hatr)\right] \cdot
\vj_{\mu }, \notag \\
\mathcal{E}_{JM} &=&i\left[ \left( \frac{J+1}{2J+1}\right)
^{1/2}F_{JJ-1}^{M}(\hatr)-\left( \frac{J}{2J+1}\right)
^{1/2}F_{JJ+1}^{M}(\hatr)\right] \cdot \vj_{c}%
+qF_{JJ}^{M}(\hatr)\cdot \vj_{\mu}.
\end{eqnarray}
With the explicit expressions of the $F_{1L}^{M}(\hatr)$,
M1 multipoles can also be written as
%be used, giving:
\begin{equation}
\mathcal{M}_{1M}=i\sqrt{\frac{3}{8\pi }}j_{1}(qr)\left[ \hatr\times
\vj_{c}\right] +\frac{iq}{\sqrt{6\pi }}\left[ j_{0}(qr)%
\vj_{\mu}-\frac{1}{2}j_{2}(qr)\left\{ \vj_{\mu}
-\hatr\left( \hatr\cdot \vj_{\mu}\right)
\right\} \right]\,.
\end{equation}
In terms of the reduced matrix elements (RMEs)~\cite{Carlson:1997qn,Pisa_nd_cap},
%One often use~\cite{Carlson:1997qn,Pisa_nd_cap} reduced matrix elements (RMEs), which are
%related to $\mathcal{X}_{JM}\equiv (\mathcal{M}_{JM},\mathcal{E}_{JM})$ by
\begin{equation}
%\left\Vert
\widetilde{\mathcal{X}}_{J}^{J_{i}J_{f}}
%\right\Vert
=\frac{\sqrt{6\pi }}{q\mu _{N}}\sqrt{4\pi }\left\langle \Psi _{b.s.}^{J_{f}}
\big\Vert \mathcal{X}_{JM}\big\Vert
\Psi _{scat}^{J_{i}}\right\rangle,
\end{equation}
where $\mathcal{X}_{JM} = (\mathcal{M}_{JM},\ \mathcal{E}_{JM})$,
%
%In terms of RMEs,
the total $nd$ capture cross section is given by
\be
\label{eq_nd_cs}
\sigma_{nd} =\frac{2}{9}\frac{\alpha }{\left( v_{rel}/c\right) }
\left( \frac{\hbar c}{2mc^{2}}\right)^{2}\left( \frac{q}{\hbar c}\right)^{3}
\sum_{J_{i}}\sum_{J=1}^{J_{i}+\frac{1}{2}}
\left(
\left|
%\left\Vert
\widetilde{\mathcal{E}}_{J}^{J_{i},\frac{1}{2}}
%\right\Vert
\right|^{2}
+\left|
%\left\Vert
\widetilde{\mathcal{M}}_{J}^{J_{i},\frac{1}{2}}
%\right\Vert
\right|^{2}
\right).
\ee
Thermal neutron capture proceeds only from doublet
$J_{i}^{\Pi }=\frac{1}{2}^{+}$ and quartet $J_{i}^{\Pi }=\frac{3}{2}^{+}$
$nd$ scattering states, since
only these two states comprise $nd$ $S$-wave asymptote and thus dominate low
energy scattering. Since final state (the triton) is
$J_{f}^{\Pi }=\frac{1}{2}^{+}$,
therefore only magnetic dipole transition elements
$m_2 \equiv
%\left\Vert
\widetilde{\mathcal{M}}_{1}^{\frac{1}{2},\frac{1}{2}}
%\right\Vert
$,
$m_4 \equiv
%\left\Vert
\widetilde{\mathcal{M}}_{1}^{\frac{3}{2},\frac{1}{2}}
%\right\Vert
$
and electric quadrupole transition element
$e_4 \equiv
%\left\Vert
\widetilde{\mathcal{E}}_{2}^{\frac{3}{2},\frac{1}{2}}
%\right\Vert
$
do not vanish.
Notice that magnetic dipole moments are purely imaginary, while electric
quadrupole moment is real.

Experimentally, in addition to capture cross section,
photon polarization
parameter $R_{c}$ can also be measured. This parameter is given by~\cite{Pisa_nd_cap}
\begin{equation}
R_{c}=\frac{1}{3}\left[
\frac{\frac{7}{2}\left\vert m_{4}\right\vert ^{2}
+ \sqrt{8}\mbox{Re}\left[ m_{2}m_{4}^{\ast }\right]
+\frac{5}{2}\left\vert
e_{4}\right\vert ^{2}
+\sqrt{24}\mbox{Im}\left[ m_{2}e_{4}^{\ast }\right]
-\sqrt{3}\mbox{Im}\left[ m_{4}e_{4}^{\ast }\right] }{
\left\vert m_{2}\right\vert^{2}
+\left\vert m_{4}\right\vert ^{2}
+\left\vert e_{4}\right\vert ^{2}}-1\right] .
\end{equation}
%where we employ shorthand notations
%$m_{2}=\left\Vert \widetilde{\mathcal{M}}_{1}^{(%
%\frac{1}{2})(\frac{1}{2})}\right\Vert$,
%$m_{4}=\left\Vert \widetilde{\mathcal{M%
%}}_{1}^{(\frac{3}{2})(\frac{1}{2})}\right\Vert $
%and
%$e_{4}=\left\Vert
%\widetilde{\mathcal{E}}_{1}^{(\frac{3}{2})(\frac{1}{2})}\right\Vert $ for RMEs.

Calculations using expression~(\ref{eq_3x_rule}) are numerically stable for
all the two- and one-body current terms except the ones entering into impulse approximation of M1 operator.
This issue has been observed and the special numerical procedure  developed
in reference~\cite{Friar_nd_cap}, we have successfully followed it.

\section{Results}

\subsection{Binding energies and scattering lengths}

In this work we have performed rigorous calculations for several qualitatively
different realistic nuclear Hamiltonians, which are based on $NN$
potentials defined both in configuration and momentum spaces.
Argonne Av18~\cite%
{POT_AV18} is an accurate local $NN$ potential
in configuration space.
Semi-realistic configuration space potential INOY has been
recently derived by Doleschall~\cite{POT_INOY}, which can
describe binding energies of three-nucleon systems
with only two-nucleon forces.
% without three-nucleon force.
ISUJ~\cite{POT_ISUJ}-- a recent revision of INOY --
%even
further improves description of
$np$ and $pp$ data and
at low energies provides solution for the
long standing ``Ay puzzle" of $N$-$d$ scattering.
We have also
tested some chiral \nlo3 potentials defined in momentum space:
Idaho group potential~\cite{POT_IN3LO} (referred to as I-N3LO), and three different
parameterizations of chiral \nlo3 potential of
Bonn-Bochum group~\cite{POT_BN3LO}.
In particular Bonn-Bochum group potentials parameterized with
set of cut-off values $\left\{ \Lambda ,\widetilde{\Lambda }\right\}
=\left\{ 450,500\right\} ,\left\{ 450,700\right\} $ and $\left\{
600,700\right\} $ MeV have been used and are referred to as B1-N3LO,
B2-N3LO and B3-N3LO respectively.

All the $NN$ potentials mentioned above describe the $NN$ data quite accurately.
%with high-accuracy experimental NN data.
And all but
%Also all except
Bonn-Bochum group potentials reproduce experimental deuteron
binding energy $B_d$ and the singlet $np$ scattering length ${}^1a_{np}$
with at
least four significative digit accuracy. Values of these observables
obtained using Bonn-Bochum group
potentials are summarized in Table~\ref{tab:Bonn_2B}.

\begin{table}[tbp]
\caption{Values for $np$ singlet scattering length and deuteron binding energy
obtained using Bonn-Bochum group potentials.}
\label{tab:Bonn_2B}%
\begin{ruledtabular}
\begin{tabular}{ccc}
Model & $^{1}$a$_{np}$ (fm) & B$_{H_{2}}$ (MeV) \\\hline
B1-N3LO & -23.60 & 2.215 \\
B2-N3LO & -23.72 & 2.218 \\
B3-N3LO & -23.64 & 2.220 \\\hline
Exp.: & -23.74 & 2.225%
\end{tabular}
\end{ruledtabular}
\end{table}

Our three-body calculations have been carried out considering isospin
breaking effects, which allow admixture of total isospin
$T=3/2$ in the wave functions.
%states in
%systems wave function.
The Argonne UIX three-nucleon interaction~\cite{POT_UIX} also has been
taken into account
in the combination with
Av18 $NN$ potential.
% has been also combined with , when calculating three-nucleon
%properties.

The relevant properties of three-body systems obtained with the adopted models
are summarized in Table~\ref{tab:b_en}.
% we present our results for binding energies and
%scattering lengths of three-nucleon system.
These values are in perfect
agreement with ones obtained by the
other groups~\cite{POT_INOY},\cite{Wit_sc_len}-\cite{Arnas_pr}.
In~\cite{Lazo_nonloc} we have already published three-nucleon properties
for INOY and Av18 models, the small difference in fourth digit of those results
compared with current ones is due to the small admixture of isospin $T=3/2$ states.
%which are considered in this work.
One should note that only INOY, ISUJ and Av18+UIX models
reproduce experimental three-nucleon binding
energies as well as neutron-deuteron doublet $(J=\frac{1}{2})$ scattering
length accurately.
Chiral potentials at \nlo3 comprise already two irreducible three-nucleon interaction
diagrams with contact terms. The strength of these contact terms may be
{\it a priori} adjusted so as to
reproduce three-nucleon binding energy and scattering length~\cite{Epelbaum:2002vt}.
In this work however only two-nucleon interaction part of \nlo3 models was considered.

\bigskip
\begin{table}[htbp]
\caption{Three-nucleon properties as calculated with different realistic
Hamiltonians. They contain: $nd$ doublet ($^{2}$a$_{nd}$) and
quartet ($^{4}$a$_{nd}$) scattering lengths in fm; bound state properties comprising binding
energy(BE), average kinetic energy ($\langle T \rangle$) in MeV's and rms radius
$r_{\rm rms}=\sqrt{\langle r^2\rangle}$  in fm.
These values are compared to other theoretical calculations and experimental
results.}
\label{tab:b_en}%
\begin{ruledtabular}
\begin{tabular}{cccccccccc}
&  & \multicolumn{2}{c}{$nd$} & \multicolumn{3}{c}{$\H3$} &
\multicolumn{3}{c}{$\He3$} \\
Hamiltonian & Ref. & $^{2}$a$_{nd}$ & $^{4}$a$_{nd}$ & BE & $\langle T\rangle$
& $r_{\rm rms}$ & BE & $%
\langle T\rangle$ &
$r_{\rm rms}$ \\\hline
Av18 & this work & 1.266 & 6.331 & 7.623 & 46.71 & 1.769 & 6.925 & 45.67 &
1.810 \\
& \cite{Wit_sc_len,Bench_the} & 1.248 & 6.346 & 7.623(2) &  &  &
6.924(1) &  &  \\
AV18+UIX & this work & 0.598 & 6.331 & 8.483 & 51.29 & 1.683 & 7.753 & 50.23
& 1.716 \\
& \cite{Wit_sc_len,Bench_the} & 0.578 & 6.347 & 8.478(2) &  &  &
7.748(2) &  &  \\
INOY & this work & 0.551 & 6.331 & 8.483 & 33.00 & 1.666 & 7.720 & 32.22 &
1.704 \\
& \cite{POT_INOY} &  &  & 8.482 &  &  & 7.718 &  &  \\
ISUJ & this work & 0.523 & 6.330 & 8.484 & 32.95 & 1.667 &  &  &
 \\
& \cite{POT_ISUJ} &  &  & 8.482 &  &  & 7.718 &  &  \\
I-N3LO & this work & 1.101 & 6.337 & 7.852 & 34.54 & 1.760 & 7.159 & 33.83 &
1.797 \\
& \cite{Arnas_be} &  &  & 7.854 &  &  &  &  &  \\
B1-N3LO & this work & 1.263 & 6.334 & 7.636 & 33.60 & 1.816 & 6.904 & 32.79
& 1.860 \\
&\cite{Arnas_pr}  &  &  & 7.64 &  &  &  &  &  \\
B2-N3LO & this work & 1.024 & 6.339 & 7.930 & 31.70 & 1.777 & 7.210 & 31.01
& 1.815 \\
&\cite{Arnas_pr}  &  &  & 7.97 &  &  &  &  &  \\
B3-N3LO & this work & 1.781 & 6.329 & 7.079 & 47.25 & 1.863 & 6.403 & 46.17
& 1.909 \\
&\cite{Arnas_pr}  &  &  & 7.09 &  &  &  &  &  \\\hline
Exp: &  & 0.65$\pm $0.04~\cite{exp_nd_scat_len} & 6.35$\pm $0.02~\cite{exp_nd_scat_len} & 8.482 & - &  & 7.718 & - &
\end{tabular}
\end{ruledtabular}
\end{table}

\subsection{Magnetic moments and thermal neutron capture}
\begin{table}[tbp]
\caption{Matrix elements calculated for magnetic moments and thermal neutron capture.
These results are obtained using INOY Hamiltonian
with $\Lambda$=700 MeV.}
%and fixing  cutoff
%value $\Lambda$=700 MeV for meson exchange currents.}
\label{tab:mat_el}%
\begin{ruledtabular}
\begin{tabular}{cccccccc}
& $\mu (\H2)$ & $\mu (\H3)$ & $\mu (\He3)$ &
$\frac{1}{i} \widetilde{\mathcal{M}}_{1}^{0,1} $ &
$\frac{1}{i} \widetilde{\mathcal{M}}_{1}^{\frac{1}{2},\frac{1}{2}}$ &
$\frac{1}{i} \widetilde{\mathcal{M}}_{1}^{\frac{3}{2},\frac{1}{2}}$ &
$\widetilde{\mathcal{E}}_{2}^{\frac{3}{2},\frac{1}{2}}$ \\\hline
LO: 1B  & 0.8593 & 2.6567 & -1.8100 & 395.5000 & -13.6196 & 13.1149 & -0.0741
\\
N$^{3}$LO: 1B & -0.0057 & -0.0199 & 0.0080 & -0.1653 & 0.4106 & 0.1048 &
0.0032 \\
NLO:  1$\pi $ & 0.0000 & 0.1515 & -0.1501 & 7.0970 & -2.5712 & -0.4289 &
0.1562 \\
N$^{3}$LO: 1$\pi C$ & -0.0029 & 0.0839 & -0.0926 & 3.1860 & -2.7674 & -0.3465
& 0.0000 \\
N$^{3}$LO: 2$\pi $ & 0.0000 & 0.0374 & -0.0362 & 1.1290 & -1.2504 & -0.1223
& -0.0019 \\
g$_{4S}$ & 0.0338 & 0.0457 & 0.0449 & 0.0000 & -0.9855 & 0.2647 & 0.0000 \\
g$_{4V}$ & 0.0000 & 0.0733 & -0.0712 & 2.3130 & -2.5179 & -0.2267 & 0.0000%
\end{tabular}
\end{ruledtabular}
\end{table}

In Table~\ref{tab:mat_el},
we present M1 RMEs
% of two- and three-nucleon systems,
obtained for INOY Hamiltonian with $\Lambda$=700 MeV,
listing the contributions from each chiral order.
Note that the one-body contribution of
the iso-scalar M1 RME, $m_2$,
%For most cases, M1 RMEs
%are dominated by the one-body contribution.
%urrent of impulse approximation.
%two-body contribution is less than one percent for two-nucleon system and
%they contribute at the level of 10\% for three-nucleon magnetic moments. Nevertheless
%The exception is
%whose LO one-body contribution
is strongly suppressed due to the pseudo-orthogonality
between initial and final wave functions.
%The other channels are dominated by the one-body contribution.
%
%The relative smallness of \nlo3 compared to NLO is however not much
The chiral convergence is however not much
illuminating, {\it i.e.}, \nlo3 contributions appear about the same size of NLO.
%As was explained in~\cite{Pastore:2008sk},
This behavior is mainly due to the accidental cancelation between two NLO contributions,
%provided by
the seagull and pion-pole diagrams~\cite{Pastore:2008sk}.
%partly due to the largeness of the resonance saturation contributions
%and  it is a result of the destructive interference of the terms
%provided by seagull and pion in-flight NLO diagrams.
%
%as much as 50\% of RMEs for thermal neutron capture on deutron from the initial doublet state is due to the
%2-body current.  This is due to the fact that the spin operator, dominating 1-body current, is
%partially suppressed: in this case initial and final nuclear states with probability
%$\sim$90\% has total spin S=1/2 and total angular momentum L=0,
%however overlap of these principal components should be small due to the orthogonality
%of wave functions.
%
%Looking at the chiral power counting, see Table~\ref{tab:NLO_an}, one can see that contribution of the LO diagrams as should governs
%any M1 matrix element. Nevertheless the total contribution of \nlo3 diagrams can be as large as contribution of
%NLO diagrams. As explained in~\cite{Pastore:2008sk} it is a result of the destructive interference of the terms
%provided by seagull and pion in-flight NLO diagrams.

\begin{table}[tbp]
\caption{Values of contact term coefficients g$_{4s}$ and g$_{4V}$ , which
are obtained by fitting magnetic moments of triton and $^{3}$He, for INOY
Hamiltonian.}
\label{tab:N3LO_gs}%
\begin{ruledtabular}
\begin{tabular}{ccc}
$\Lambda$ (MeV) & g$_{4s}$ & g$_{4V}$ \\\hline
500 & 0.2747 & 1.8746 \\
%600 & 0.2475 & 1.1954 \\
700 & 0.2313 & 0.8021 \\
%800 & 0.2122 & 0.5820 \\
900 & 0.1997 & 0.4613%
\end{tabular}
\end{ruledtabular}
\end{table}

As explained,
M1 currents contain two non-derivative contact-terms at \nlo3.
Since the coefficients of them,
%LECs associated with these terms,
$g_{4S}$ and $g_{4V}$, cannot be determined from the underlaying theory yet,
we fit these constants by
requiring that magnetic moments of $^3$H and $^3$He are correctly reproduced.
The resulting values obtained with INOY potential
are given in Table~\ref{tab:N3LO_gs}.
We remark that $g_{4S}$ and $g_{4V}$ depend on cutoff $\Lambda$ as well as on
particular choice of nuclear Hamiltonian.
% in use.
%\footnote{
%({\bf TSP: Let's remove following part, which is not quite necessary,
%and find good english is difficult.})
%In our preceding study~\cite{A23} we have fitted these
%parameters in order to reproduce two-nucleon M1 properties
%-- deuteron magnetic moments and thermal $np$ capture cross section.
%Taking the tri-nucleon magnetic moments as reference data
%has the advantage that
%these quantities are more sensitive
%to $g_{4S}$ and $g_{4V}$ values,
%with little error bar in experimental data
%($np$ thermal capture cross section used as reference data in~\cite{A23}
%is determined with only 0.2\% accuracy).
%}

\begin{table}[htbp]
\caption{
Dependence of M1 observables for two and three-nucleon systems on cutoff
value $\Lambda$. These results are obtained using INOY Hamiltonian.}
\label{tab:NLO_an}%
\begin{ruledtabular}
\begin{tabular}{ccccccc}
\multicolumn{7}{c}{LO} \\
$\Lambda$ (MeV) & $\mu (\H2)$ & $\mu (\H3)$ & $\mu (\He3)$ &
$\sigma_{np}$ (mb) & $\sigma_{nd}$ (mb) & $R_{c}$ \\\hline
-& 0.8593 & 2.657 & -1.810 & 309.7 & 0.2785 & -0.2369 \\\hline\hline
\multicolumn{7}{c}{NLO} \\
$\Lambda$ (MeV) & $\mu (\H2)$ & $\mu (\H3)$ & $\mu (\He3)$ & $\sigma
_{np}$ (mb) & $\sigma _{nd}$ (mb) & $R_{c}$ \\\hline
500 & 0.8593 & 2.760 & -1.913 & 318.7 & 0.2972 & -0.3026 \\
%600 & 0.8593 & 2.789 & -1.941 & 320.0 & 0.3160 & -0.3344 \\
700 & 0.8593 & 2.808 & -1.960 & 320.9 & 0.3296 & -0.3538 \\
%800 & 0.8593 & 2.821 & -1.972 & 321.5 & 0.3404 & -0.3670 \\
900 & 0.8593 & 2.829 & -1.980 & 321.9 & 0.3480 & -0.3753 \\\hline\hline
\multicolumn{7}{c}{\nlo3 without contact term} \\
$\Lambda$ (MeV) & $\mu (\H2)$ & $\mu (\H3)$ & $\mu (\He3)$ & $\sigma
_{np}$ (mb) & $\sigma _{nd}$ (mb) & $R_{c}$ \\\hline
500 & 0.8499 & 2.836 & -2.011 & 324.1 & 0.3612 & -0.3896 \\
%600 & 0.8503 & 2.881 & -2.054 & 326.2 & 0.3983 & -0.4205 \\
700 & 0.8507 & 2.910 & -2.081 & 327.5 & 0.4237 & -0.4366 \\
%800 & 0.8511 & 2.927 & -2.096 & 328.3 & 0.4416 & -0.4456 \\
900 & 0.8515 & 2.937 & -2.105 & 328.8 & 0.4526 & -0.4504 \\\hline\hline
\multicolumn{7}{c}{\nlo3} \\
$\Lambda$ (MeV) & $\mu (\H2)$ & $\mu (\H3)$ & $\mu (\He3)$ & $\sigma
_{np}$ (mb) & $\sigma _{nd}$ (mb) & $R_{c}$ \\\hline
500 & 0.8584 & 2.9790 & -2.1276 & 330.9 & 0.5012 & -0.4659 \\
%600 & 0.8584 & 2.9790 & -2.1276 & 330.7 & 0.4974 & -0.4653 \\
700 & 0.8585 & 2.9790 & -2.1276 & 330.5 & 0.4946 & -0.4649 \\
%800 & 0.8583 & 2.9790 & -2.1276 & 330.5 & 0.4951 & -0.4649 \\
900 & 0.8583 & 2.9790 & -2.1276 & 330.4 & 0.4959 & -0.4650 \\\hline
Exp.: & 0.8574 & 2.9790 & -2.1276 & 332.6 $\pm $0.7 & 0.508$\pm $0.015 &
-0.420$\pm $0.030\\
\end{tabular}
\end{ruledtabular}
\end{table}

Table~\ref{tab:NLO_an} shows the cutoff dependence of our results.
One-body contributions are cutoff independent by their construction.
NLO results bring sizable cutoff-dependence, indicating that
some important pieces are omitted at this level.
As is indicated in the table,
going \nlo3 but without taking the CTs does not help in resolving the situation.
It is only after the CTs taken into account that the results become almost independent
of the cutoff,
which implies that the CTs are quite effective in renormalizing away the
details residing in the short-range region.

%NLO terms brings in strong cutoff dependence, which
%reflects importance of high-energy (or short-range) physics we neglect at this order.  If contact diagrams are neglected
%\nlo3 terms brings even stronger cutoff dependence. On the other hand MEEFT regularization procedure,
%which permits to determine values of LECs  $g_{4S}$ and $g_{4V}$, drastically reduce cutoff dependence for
%all M1 observables we have calculated in this study. Cutoff dependence falls below 1\% level compared
%with total 2-body current contribution. This enables one make predictions with three significative digits
%for two-body system, whereas thermal neutron capture on deuteron is calculated with 0.5\% accuracy,
%see Table~\ref{tab:final_res}, where calculated M1 observables are summarized for different nuclear
%Hamiltonians.

%xxxxxxxxxxxxxxxxxxxxxxxxxxxxxxxxxxxxxxx
%In Table~\ref{tab:final_res} one can see that theoretical predictions
%by any considered Hamiltonian
%fail to reproduce experimentally measured values. Thermal
%np capture cross section is systematically underestimated. Deuteron magnetic moment
%is overestimated for all except underestimating B3-N3LO Hamiltonian and
%three Hamiltonians (AV18,I-N3LO and B1-N3LO) providing exact value. Nevertheless
%experiment-theory missmatch  for 2-body observables is rather small and accounts
%less than 30\% of \nlo3 contributions. As mentioned in~\cite{A23} one can
%reasonably expect that \nlo4 effects remedy this discrepancy.

\begin{table}[hbp]
\caption{Predictions for the deuterons magnetic moment and the observables of
the thermal neutron capture on protons and deuterons. These calculations have
been realized by fixing contact terms of the meson exchange current in order
to reproduce magnetic moments of the triton and $^{3}$He. These values turns
to be insensitive to the cut-off parameter in the interval
$\Lambda=(500,900)$ MeV; if however variation was larger than one affecting the
fourth significant digit it is given in parentheses.  The constructed AV18+UIX* model
gives $^2a_{nd}$=0.623 fm, $^4a_{nd}$=6.331 fm and BE($\He3$)=7.718 MeV; the I-N3LO+UIX**
results are $^2a_{nd}$=0.634 fm, $^4a_{nd}$=6.339 fm and BE($\He3$)=7.737 MeV. Both these
models are adjusted to reproduce experimental triton binding energy of BE($\H3$)=8.482 MeV }
\label{tab:final_res}%
\begin{ruledtabular}
\begin{tabular}{ccccc} %|cccc}
Model    & $\mu(\H2) $ & $\sigma_{np}$ $(mb)$   & $\sigma_{nd}$ $(mb)$ & $R_{c} $\\\hline
%& ${}^2a_{nd}$ & ${}^4a_{nd}$ & BE($\H3$) & BE($\He3$)
AV18     & 0.8575& 331.9(1)          & 0.680(3) & -0.435\\
%& 1.266  &  6.331 & 7.623 & 6.925
AV18+UIX & 0.8604& 330.6(2)          & 0.478(3) & -0.458\\
%& 0.598  &  6.331 & 8.483 & 7.753 \\
INOY     & 0.8585&  330.6(2)         & 0.498(3) & -0.465\\
%& 0.551  &  6.331 & 8.483 & 7.720 \\
ISUJ     & 0.8585&  331.1(2)         & 0.501(2) & -0.466\\
%& 0.523  &  6.330 & 8.484 & 7.718 \\
I-N3LO   & 0.8574 & 330.4(3)          & 0.626(2) & -0.441\\
%&1.101 & 6.337 & 7.852 & 7.159\\
B1-N3LO  & 0.8577 & 328.7(6)          & 0.688(4) & -0.438(1)\\
%& 1.263 & 6.334 & 7.636 & 6.904\\
B2-N3LO  & 0.8588  &331.0(4)           & 0.609(4) & -0.448(1)\\
%&1.024  & 6.339 & 7.930 & 7.210\\
B3-N3LO  & 0.8549  &330.9(7)          & 0.879(8) & -0.411(2)\\
%& 1.781 & 6.329 & 7.079 & 6.403 \\\hline
%$\langle$ extrapolated $\rangle$  & -  & -          & 0.490(8) & -0.462(3)
%& - & - & 8.482 & - \\
\hline
AV18+UIX* & 0.8614(1) & 330.9(3)     & 0.476(2) & -0.457(1)\\
%& 0.623  &  6.331 & 8.482 & 7.718 \\
I-N3LO+UIX** & 0.8590(1) & 329.7(3)          & 0.477(3) & -0.468(1)\\ \hline
%& 0.634  &  6.339 & 8.482 & 7.737 \\
Exp.:    & 0.8574 &332.6 $\pm $0.7~\cite{np_capt_exper}    & 0.508$\pm $0.015~\cite{nd_capt_exper}& -0.420$\pm $0.030~\cite{rc_capt_exper}\\
%& 0.65 $\pm$ 0.04 & 6.35 $\pm$ 0.02 & 8.482 & 7.718
\end{tabular}
\end{ruledtabular}
\end{table}

\begin{figure}[h]
\includegraphics[width=8.25cm]{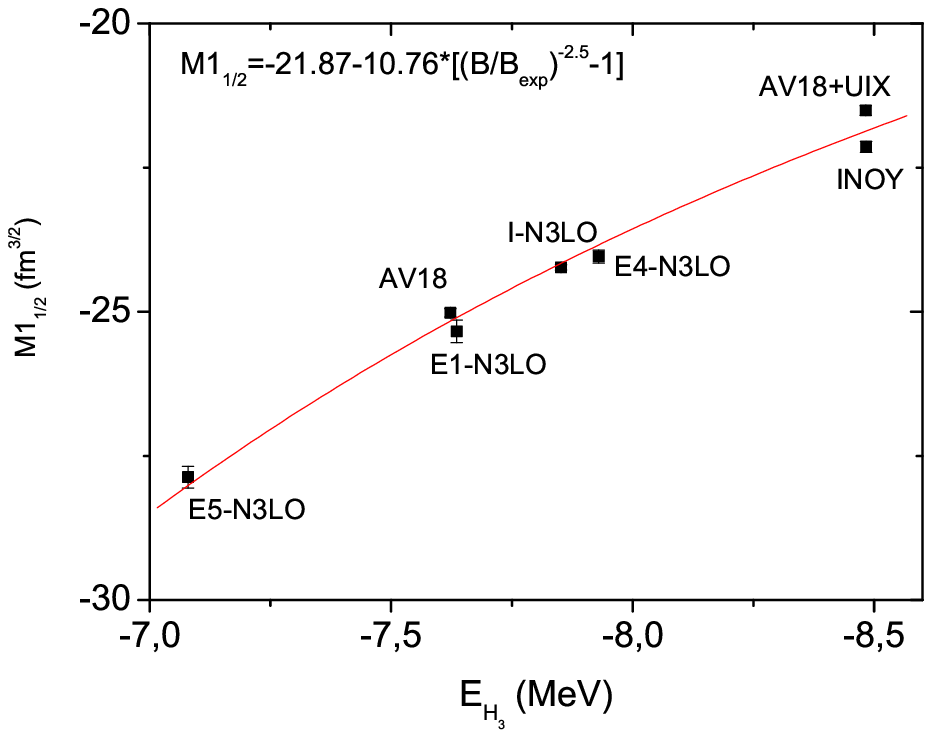} \hskip 1cm %
\includegraphics[width=8.25cm]{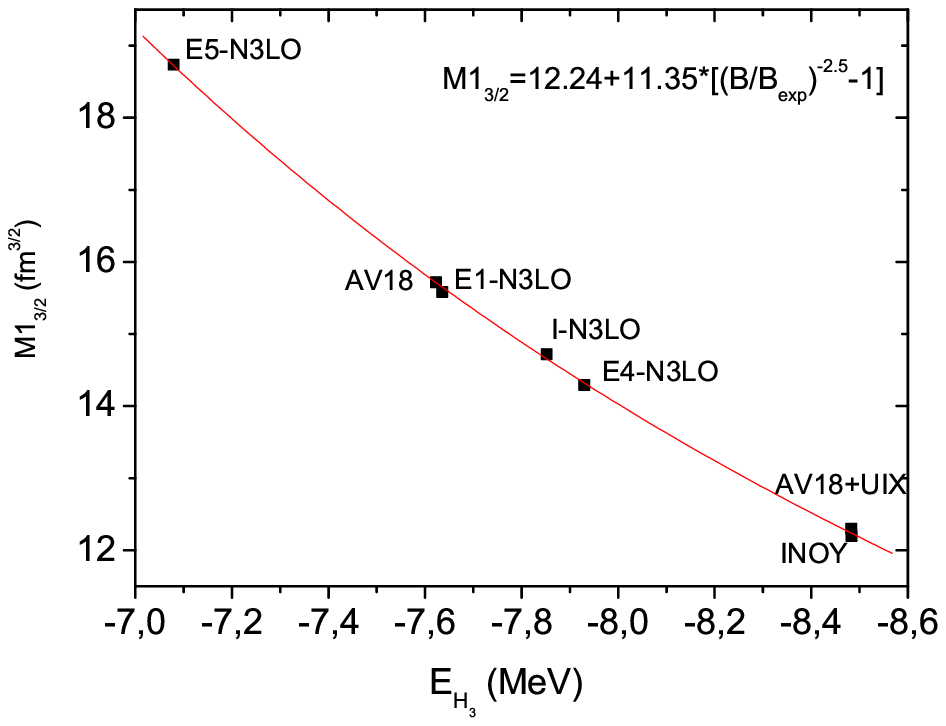}
\caption{Radiative capture of thermal neutron by deutron: correlation of M1
doublet and quartet RME's with triton binding energy.}
\label{fig:detcomp}
\end{figure}

Results with varying model Hamiltonian are given in Table~\ref{tab:final_res},
with some relevant low-energy properties of the potentials.
From the table, one observes that
$\mu(\H2)$ and $\sigma_{np}$ are rather insensitive,
$R_c$ is moderately sensitive
and
the $nd$ capture cross section, $\sigma_{nd}$,
is highly sensitive on the model Hamiltonian.
To understand the sensitivity,
let us consider the model-dependence of the effective-range parameters (ERPs),
which govern the long-range part of the RMEs.
The most important ERPs are the binding energies
and the scattering lengths,
which should strongly influence M1 RMEs through the
coupling of the long-range parts of the three-nucleon wave functions.
And indeed,
as shown in Fig.~\ref{fig:detcomp},
the M1 RMEs
%, $m_2$ and $m_4$,
are strongly
correlated with the triton binding energy $B_3$.
The correlation is found to be almost perfect for $m_4$,
while with some fluctuation for $m_2$.
These behavior can be explained with simple arguments:
let us first concentrate on the quartet RME, $m_4$.
In spin-quartet states, Pauli principle inhibits three nucleons
from gathering altogether, and thus observables are insensitive to short-range
part of three-nucleon interaction.
As a result,
the $nd$ quartet scattering length $^4a_{nd}$ has little model-dependence;
all the models considered here reproduce $^4a_{nd}$
in excellent agreement with the experimental data.
This explains the perfect correlation of $m_4$ with $B_3$.
%
%As a result, only triton's ERPs enter for $m_4$,
%and the observed perfect correlation of $m_4$ with
%the triton BE $B_3$
%can be understood.
%
%Now let us consider spin-doublet case,
%where the situation is rather complicate, %As has been demonstrated in the pionless EFT,
%since
On the contrary,
spin-doublet states are free from the exclusion principle and sensitive to the
short-range three-nucleon interaction.
%This may also be inferred by noticing that there is
%As a result,
%there is sizable model-dependence on the doublet
This makes the scattering length $^2a_{nd}$
%has large model-dependence,
largely model-dependent,
see Table~\ref{tab:final_res},
and we might expect that $m_2$ depends
not only on $B_3$ but also on $^2a_{nd}$.
However $^2a_{nd}$ and $B_3$ are correlated,
which is known in terms of the Phillips line~\cite{phillips}.
The correlation of $^2a_{nd}$ with $B_3$ is not perfect,
showing small deviations from the Phillips line.
These arguments are in good accordance with what we observe in
%the right panel of
Fig.~\ref{fig:detcomp},
which shows the correlation of $m_2$ with respect to $B_3$ with some scatters.

%For example,
%the relation of the two can be read using effective range expansion,
%$\gamma_3 = {^2a_{nd}}^{-1}  + \cdots$
%where
%the ellipses denotes higher order terms in the effective range expansion,
%$\gamma_3=\sqrt{2 m_{nd} B_3}$ and $m_{nd}$ is the reduced mass of $nd$ systems.
%

%The relation is not perfect due to higher order terms.
%Indeed
%EFT allows two different contact three-nucleon interactions
%at the first non-vanishing order,
%and one can fix the coefficients of the TNIs to have
%experimental value of $B_3$ and $^2a_{nd}$.
%
%Relaying on these observations,
%it will be reasonable to expect a correlation of $m_2$ with $B_3$
%with some residual model-dependence,
%which is in accordance with
%Fig.~\ref{fig:detcomp}.

%The correlation might be understood by recalling that
%Since the long-range parts of the RMEs are governed by the effective-range parameters (ERPs)
%like $B_3$,

%
%
%such as the binding energy, the scattering length
%and the effective range
%of the initial and final states.

%
For noble two-body processes,
effective range expansion technique often allows us
even algebraic relation of the RMEs in terms of ERPs,
see, for example, Refs.~\cite{ApJ} for the
Gamow-Teller matrix element of the $p+p\to d+e^++\nu$ process.
The problem at hand is however too complicate to allow such
a mathematical rigor,
and we will limit ourselves to an empirical curve fitting.
We take the trial function as
\be
%m_n^{(i)} \simeq \phi_n(B_3^{(i)})= m_n^0 + t_n \left[(B_3^{(i)}/B_3^{\rm exp})^\nu-1\right],
m_n^{(i)} \simeq \phi_n(B_3^{(i)})
\ee
with
\be
\phi_n(B_3)= m_n^0 + t_n \left[(B_3/B_3^{\rm exp})^\nu-1\right],
\ee
where the superscript $i$ is the model index; that is,
$m_n^{(i)}$ ($n=2,4)$ and $B_3^{(i)}$
stands for the RMEs and $^3H$ BE obtained with the $i$-th model potential, respectively.
Varying the value of $\nu$, values of $m_n^0$ and $t_n$ are searched by a chi-square fit.
%and $\nu$ are free parameter,
%$n=(2,\,4)$,
The resulting chi-square is found to be parabola shape with minimum at around $\nu=-2.5$.
The solution with $\nu=-2.5$ is
\be
\phi_2(B_3) &=& (-21.87 \pm 0.24) - 10.76 \left[(B_3/B_3^{\rm exp})^{-2.5}-1\right],
\nonumber \\
\phi_4(B_3) &=& (12.24 \pm 0.05) + 11.35 \left[(B_3/B_3^{\rm exp})^{-2.5}-1\right].
\label{phin}\ee
The solution is drawn in solid line in the figure.
The above curve fitting procedure turns out to be quite robust.
For example,
the curves and the values of $m_n^0$ with $\nu=-1.5$ are almost the same as those with $\nu=-2.5$.
Even if we try a simple-minded linear fit, $\nu=1$,
we have $\phi_2(B_3)=-21.73 + 33.65 (x_3-1)$ and
$\phi_4(B_3)=12.14 - 35.69 (x_3-1)$,
$x_3\equiv E_{\rm H3}/E_{\rm H3}^{\rm exp}$.
Thus
the values of $\phi_n(B_3^{\rm exp})=m_n^0$ are quite insensitive to the fitting parameter $\nu$.
Furthermore,
with the resulting values of $\phi_n(B_3^{\rm exp})=m_n^0$,
we have $R_c$=-0.462$\pm$0.03 and
$\sigma_{nd}$=0.490$\pm$0.008 mb,
which are close to the experimental data.
Therefore one can conclude that
the observed strong model-dependence in M1 properties of three-body systems can be
traced to the different model predictions of $B_3$,
and that,
once we have correct $B_3$,
the theory predictions should be very close
to the experimental data
with little model-dependence.
%
%Correlation behavior of  eq.(\ref{phin})
%thus implies reproducing the trinucleon binding energy correctly is
%crucial to reproduce $nd$ thermal capture cross section. In fact, if one
%considers experimental q-value in  eq.(\ref{eq_nd_cs}) then nd cross section is overestimated
%for trinucleon binding energy underestimating models. If one takes calculated q-value
%then nd cross section becomes underestimated for trinucleon binding energy underestimating
%models.  Photon circular polarization parameter $R_c$ has no direct dependence on trinucleon
%binding energy, however one can still see some correlation. This parameter tends to
%decrease with trinucleon binding energy.

%By comparing different model predictions
% one can see that models capability
%to reproduce trinucleon binding energy sensibly improves its potential to reproduce
%nd capture cross section, while photon polarization parameter $R_c$ is more or less stable.

We have also tried to adjust the nuclear potentials to have correct ERPs.
As mentioned, $B_{3}$ and $^2a_{nd}$ are the relevant ERPs.
But since the two ERPs are strongly correlated to each other,
simultaneous reproduction of both is rather tricky.
This correlation in particular strong due to on-shell $NN$ interaction
part, nevertheless three-nucleon interaction can break it.
%It is known that the correlation can be broken by the TNIs.
%
Note that UIX TNI potential consists of two terms.
In our calculation,
%
%One should note that if INOY, ISUJ and AV18+UIX models reproduce well trinucleon binding energies,
%they slightly underestimate nd doublet scattering length. As noted above $B_{3}$ and
% $^2a_{nd}$ correlate.  To see effect of $^2a_{nd}$
% on nd capture observables
we have readjusted the parameters of those terms
to reproduce $B_3$ and $^2a_{nd}$ simultaneously
with the Av18 and I-N3LO $NN$ potential.
We refer respectively the resulting Hamiltonian as
AV18+UIX* and I-N3LO+UIX**.
%the UIX 3NF potential arameters for AV18 and I-N3LO models, by
% constructing nuclear Hamiltonians AV18+UIX* and I-N3LO+UIX**,
%which are able to reproduce $B_3N$ and $^2a_{nd}$ simultaniously.
In addition some charge
dependence has been added to UIX*, permitting Av18+UIX* to reproduce also $^3$He
binding energy.
The corresponding results are given in the bottom lines of the Table~\ref{tab:final_res}.
%For an easy and comprehensive comparison,
%we have also listed the relative errors in Table~\ref{tab:errors}.
%
The most important observation to be made is that, while the results of
Av18, Av18+UIX and I-N3LO differ dramatically,
the modified Hamiltonians Av18+UIX* and I-N3LO+UIX**
give us almost identical results,
which confirms the argument that
our theory predictions are model-independent
once the ERPs are correctly encoded.
The resulting $\sigma_{nd}$ and $R_c$ are close to the experimental data,
but with discrepancy of about two sigmas of the data.
%
%Another thing we observe in the table is that,
%while the accuracy of the theory predictions $\mu_{H_2}$ and $\sigma_{np}$ should be regarded
%quite satisfactory for all the considered cases,
%%are very small,
%they become worse by correcting the TNIs.
%These issues will be discussed in following section.
%\footnote{
%{\bf YHS: I don't think this matter is properly discussed.
%Thus, we can remove
%this sentence or include some comments.
%Actually, changes from correcting TNIs are small and may be
%thought as the same order of neglected higher order corrections.}
%}
%one can see that
%these modifications had only minor effect on $nd$ capture observables.

Before closing this section,
we would like to make comparison with other calculations for the processes
considered in this paper.
Viviani et. al.~\cite{viviani,vv} has calculated the M1 properties of $A=2,\,3$ systems
with the currents deduced from the adopted nuclear potentials
using gauge invariance, adding model-dependent pieces for those part that
are not fixed by the gauge symmetry alone.
Their results have some variations depending on the adopted potentials~\cite{viviani} and
the details of the treatment of the
currents. Without model-dependent current part capture cross section is underestimated $\sigma_{nd} = (0.418 \sim 0.462)$ mb,
nevertheless one gets $R_c = -(0.429 \sim 0.446)$ quite close to experimental value~\cite{vv}.
Model-dependent currents enable to  reproduce experimental cross section, however
the photon polarization parameter $R_c = -(0.469)$ becomes
larger than the experimental data.
% too large like in our study.
Currents related with three-nucleon force further increase capture cross section
and photon polarization parameter.
A very similar to ours calculation has been recently performed by
Pastore et. al.~\cite{pastore},
in which electromagnetic current operators have been obtained
up-to \nlo3 within EFT framework.
%has performed a similar calculation
%recently~\cite{pastore}.
$\Delta$-isobar as well as pions and nucleons
are treated as pertinent degrees of freedom.
And they have applied the currents up-to \nlo2 to $A=2$ and $A=3$ systems.
To this order, the CT terms -- that play a crucial role in removing the
model-dependence at short-range physics -- do not appear,
and they have observed a large cutoff dependence
with a substantial under-predictions for $\sigma_{nd}$ and $R_c$,
$\sigma_{nd}= (0.450 \sim 0.315)$ mb and $R_c = -(0.437 \sim 0.331)$
for the momentum cutoff $\Lambda= (500 \sim 800)$ MeV.
%On finalizing our work, we have
%
We also acknowledge that,
using the so-called pionless EFT approach,
% has scored
%quite an impressive successes for $\sigma_{nd}$ and $R_c$.
%On the other hand, there have been impressive successes in the
%pionless effective theory approach.
Sadeghi et. al.~\cite{sadeghi} have performed up-to \nlo2 (in their counting scheme)
calculation for the $\sigma_{nd}$ and $R_c$,
achieving a perfect agreement with the data.
%\footnote{
%{\bf (YHS: We need to add more comments.)} We believe their success
%can be understood partly that they fitted some of their LECs to reproduce ERPs.
%As we showed in this work, once the ERPs are properly incorporated in the theory,
%EFT can give good agreement with experiments.
%However, calculation of Magnetic moment of 3-body
%using pionless EFT is not available yet.
%}
In their calculations,
the $np$ cross section as well as the $nd$ scattering lengths and the
binding energies (of $A=2$ and $A=3$ systems) are taken as
inputs needed to fix their parameters,
the magnetic moments
have not been considered.
Since magnetic moments are sensitive to the $D$-wave components of the wave functions,
it may not be trivial to have accurate theory predictions for the magnetic moments
using the pionless EFT.
A further study in this issue will be extremely interesting.
%it might be interesting to evaluate the magnetic moments pionless EFT
%and to make a comparison with ours.
%extend their
%evaluation of the magnetic moments using pionless EFT will
%And it might also be interesting to study
%the magnetic moments using their approach,
%have not been considered.
%It remains to be seen

\section{Discussions}

%\subsection{Three-body current contributions}

The most natural candidate for the remaining small discrepancy might be the
three-body current contributions, which are \nlo4.
It is not difficult to notice that
the leading three-body contributions are suppressed for both
M1 currents and the nuclear potentials,
for exactly the same reason.
Furthermore, the soft one-pion-exchange appears as the leading two-body contributions
for both of them.
Thus we expect that
%while the soft one-pion-exchange appears as the leading two-body contributions
%.
%By drawing some simple diagrams,
%comparing the relevant diagrams,
%one can easily find the
the ratio of the three-body contribution to the two-body contribution
is the same order for the M1 RMEs and the nuclear potentials,
%${\cal M}_{\rm 3B}/ {\cal M}_{\rm 2B}$ is of the same order with the
%ratio of the three-nucleon potential compared to the two-body potential,
\be
\frac{{\cal M}_{\rm 3B}}{ {\cal M}_{\rm 2B}}
\sim \frac{\langle V \rangle_{\rm 3B}}{\langle V \rangle_{\rm 2B}}
\sim (0.05 \sim 0.1).
\label{naive}\ee
Since the TNIs play a crucial role in reproducing the ERPs of three-body systems accurately,
we may naively guess that the same will also be true for the relation
between three-body currents and the M1 properties.
More quantitatively, eq.(\ref{naive}) with Table~III tells us that
the three-body current contribution will be about $(2 \sim 4)\ \%$
for $m_2$ and $m_4$,
which is just the needed size to remove the discrepancy of $\sigma_{nd}$ and
$R_c$. The same has been demonstrated by Viviani et al.~\cite{vv}, where
3-nucleon currents have let to increase neutron thermal capture
cross section by 0.033 mb.
In this regard, taking into account the three-body current contribution
-- while ignoring other pieces of \nlo4 for simplicity --
might be extremely interesting.

\section{Conclusion}

In this paper M1 properties, comprising magnetic moments and radiative capture of thermal
neutron observables, are studied in two- and three-nucleon systems. We utilize
meson exchange current derived up to N$^3$LO using heavy baryon chiral
perturbation theory a la Weinberg. At
N3LO, two unknown parameters, $g_{4s}$ and $g_{4v}$, enter as the coefficients of contact terms.
Following the MEEFT strategy, we have fixed them by imposing the renormalization
condition that the magnetic moments of tritium and $^3$He are reproduced. Then
we analyze the predictions for other M1 properties: magnetic moment of deuteron, as well
as observables of the thermal neutron capture on proton and deuteron.
Analysis comprise
several qualitatively different realistic nuclear Hamiltonians, which allows
us to judge on the model dependence of our results.
We obtain stable, cut-off independent results, which reconfirms efficiency of MEEFT procedure.
Model predictions for two-body observables (deuteron magnetic moment and thermal $np$ capture cross
section) scatter closely around the experimentally measured values.
%Nevertheless we observe that
%this agreement is worse for the models containing three-nucleon interaction.

Radiative capture cross section of thermal neutron on deuterons varies a lot from one Hamiltonian to
the other. We have demonstrated that this variation is mostly due to the correlation of the
capture cross section with a model predicted three-nucleon binding energy. By fixing
three-nucleon binding energy to the experimental value one can reduce model dependence below 2\%
level and obtain model-independent predictions for thermal capture cross section
$\sigma_{nd} =0.490\pm 0.008\ \mbox{mb}$ and photon polarization parameter
$R_c =-0.462 \pm 0.03$.
Within these model-dependent error bars capture cross section agrees with experimentally
measured value $0.508\pm 0.015\ \mbox{mb}$~\cite{nd_capt_exper}.
However photon polarization parameter $R_c$ is obtained
slightly too large, like in other studies based on realistic nuclear Hamiltonians and
currents~\cite{vv}. The remaining discrepancy is comparable in size with higher order
terms of the EFT, which have been neglected here. We believe that in particular three-nucleon currents,
which first appear at  N$^4$LO in our power counting scheme, should be important.

\section*{Acknowledgement}
The work of TSP was supported in part by KOSEF Basic Research Program
with the grant No. R01-2006-10912-0. The numerical calculations have been performed
at IDRIS (CNRS, France). We thank the staff members of the IDRIS computer center
for their constant help.
TSP is grateful to Prof. Daniel Phillips and Prof. Seung-Woo Hong for valuable discussions.
Last but not least we would like to thank
Prof. Mannque Rho for his kindness accepting to read the manuscript prior to its publication and
giving us important comments.

\newpage
%=============================================================================

\bibliographystyle{plain}
\bibliography{apssamp}

\end{document}